\documentclass{ws-ijmpa}
\def\1{\c{c}}
\def\2{\c{C}}
\def\3{\u{g}}
\def\4{\u{G}}
\def\5{{\i}}
\def\6{\.{I}}
\def\7{\"{o}}
\def\8{\"{O}}
\def\9{\c{s}}
\def\0{\c{S}}
\def\*{\"{u}}
\def\_{\"{U}}

\begin{document}

\markboth{ K. Karaca and S. Bay\5n}
{An Open Inflationary Model for Dimensional Reduction}

%
\catchline{}{}{}{}{}
%

\title{AN OPEN INFLATIONARY MODEL FOR DIMENSIONAL REDUCTION 
AND ITS EFFECTS ON THE OBSERVABLE PARAMETERS OF THE 
UNIVERSE}

\author{\footnotesize KORAY KARACA and SEL\2UK BAYIN} \address{Department of Physics, Middle East Technical University, 06531, Ankara, Turkey korayk@newton.physics.metu.edu.tr, bayin@metu.edu.tr}

\maketitle


\begin{abstract}
Assuming that higher dimensions existed in the early stages of the 
universe where the evolution was inflationary, we construct an open, 
singularity-free, spatially homogeneous and isotropic cosmological model 
to study the effects of dimensional reduction that may have taken place 
during the early stages of the universe. We consider dimensional reduction 
to take place in a stepwise manner and interpret each step as a phase transition. 
By imposing suitable boundary conditions we trace their effects on the present day 
parameters of the universe.

\end{abstract}

\section{Introduction}	

During the past two decades, several closed cosmological models aiming the singularity-free 
description of the universe have been proposed.\cite{Israelit,Starkovich,Bayin} In these 
works, the universe was modeled as a closed Friedmann-Lemaitre-Robertson-Walker space-time, 
bouncing from a Planck mass and radius state with inflation, connecting to the standard model of 
radiation and then to the era of matter dominance. Although they correctly described the evolution
of the universe, these models could not predicted the present value of the Hubble parameter
within the observed range. The approach developed by Bay\5n, Cooperstock and Faraoni\cite{Bayin} was 
applied to open geometry by Karaca and Bay\5n\cite{Karaca} and it was shown that a singularity-free 
cosmological model which complies with the recent measurements is possible for open geometry. In this model, 
since the initially static condition $(\dot{a}(0)=0)$ used in the closed models can no longer 
be used in the open universe case, the universe starts with an initial expansion rate, i.e., 
$\dot{a}(0)\neq 0$. Hence, this model is a two-parameter universe model in which one of the parameters 
determines the strength of the initial vacuum dominance and the other corresponds to the initial expansion rate.

It is well known that some theories require the existence of higher dimensions in the early
stages of the universe. In this paper, we investigate the possibility that higher dimensions
might have existed during the inflationary prematter phase of the universe and explore its 
observable consequences. As in the previous papers\cite{Israelit,Starkovich,Bayin,Karaca}
we adopt the Gliner\cite{Gliner}-Markov\cite{Markov} picture of Planck limits to the physical quantities.

\section{Description of the Model}

In our model, universe starts its journey as a $D$ $(D>3)$ dimensional open Friedmann model
($D$ denotes the number of initial space dimensions). At $t=0$ density of the universe is taken as 
the Planck density. Existence of higher dimensions and their reduction to three dimensions 
are assumed to take place during the inflationary prematter phase where the equation of state 
is given as
\begin{equation}
P_{D}=\left( \gamma_{D}-1\right) \rho_{D}.
\end{equation}
Here, $\gamma_{D}$ is assumed to be a positive, small and dimensionless constant that characterizes
the prematter in $D$ dimensions, and $P_{D}$ and $\rho_{D}$ represent the pressure and energy density in $D$
dimensions, respectively.

During this phase the universe expands isentropically while its temperature increases. This expansion 
continues until the maximum allowed temperature, i.e., the Planck temperature of that dimension
$(T_{pl,D})$ is reached. At this point, we postulate that a phase transition to a lower dimension
takes place, where the Planck temperature is higher. Hence, the universe finds more room for further
inflation. This process continues until we reach $D=3$. At this point, either the
last reduction may carry the universe directly to the standard radiation era 
$(P=\frac{1}{3}\rho)$ or prematter era may continue once more in the usual
three-dimensional space and then comes the radiation era. In either case,
the universe is eventually carried to the era of matter dominance $(P=0)$. In this work, we will consider
the second alternative.

Naturally, neither the dimensional reduction takes place at an instant nor we
could expect the universe to remain homogeneous and isotropic during dimensional reduction. However, we 
expect the duration of this process to be very short compared to the times that the universe spends 
in constant $D$ eras where
it is expanding as a $D$ dimensional Friedmann model. Details of what happens during the dimensional
reduction phase is beyond the scope of the present work. However, by imposing suitable boundary 
conditions at the transition points we could trace the effects of extra dimensions on the currently
observable parameters of the universe, i.e., Hubble constant, age and density.

\section{Boundary Conditions and Solutions for the Scale Factor}
We assume open, homogeneous and isotropic $(1+D)$ dimensional spacetimes where
the metric is given as:
\begin{eqnarray}
ds^{2}=c^{2}dt^{2}-a_{D}^{2}(t)[d\chi^{2}+\sinh^{2}\chi d\theta _{1}^{2}+\sinh
^{2}\chi \sin^{2}\theta _{1} d\theta _{2}^{2}+.....\\
.....+\sinh ^{2}\chi \sin ^{2}\theta _{1}\sin ^{2}\theta _{2}.....\sin ^{2}\theta _{D-2}d\theta
_{D-1}^{2}],\nonumber 
\end{eqnarray}
where $a_{D}(t)$ represents the scale factor of the universe, $t$ is the comoving
time,  $0\leq \theta _{n}\leq \pi ,$ $0\leq \theta _{D-1}\leq 2\pi
,0\leq \chi \leq \infty$ and $n=1,2,.........,D-2.$ We also assume that the
universe is filled with a perfect fluid which is represented by a
stress-energy tensor having the following nonvanishing components:
\begin{equation}
T_{0}^{0}=\rho _{D},\hskip 0.3 cm T_{1}^{1}=T_{2}^{2}=.....=T_{D}^{D}=-P_{D}. \label{3}
\end{equation}
Einstein's gravitational field equations in $(1+D)$ dimensional
space-time are (see, e.g., Ref. 7, pp. 470-482)

\begin{equation}
\frac{D(D-1)}{2a_{D}^{2}}\left[ \left( \frac{\dot{a}_{D}}{c}\right) ^{2}-1%
\right] =2\frac{A_{D}G_{D}}{c_{D}^{4}}\rho _{D},  \label{2}
\end{equation}

\begin{equation}
\left( 1-D\right) \left( \frac{\ddot{a}_{D}}{c_{D}^{2}a_{D}}\right) -\frac{%
\left( D-1\right) \left( D-2\right) }{2a_{D}^{2}}\left[ \left( \frac{\dot{a}%
_{D}}{c_{D}}\right) ^{2}-1\right] =2\frac{A_{D}G_{D}}{c_{D}^{4}}P_{D},
\label{3}
\end{equation}
where the subscript $D$ denotes the value of that quantity in $D$ dimensions, 
$A_{D}$ is the solid angle subtended by a sphere in a $D$ dimensional
space $(A_{3}=4\pi ,$ $A_{4}=2\pi ^{2},$ $A_{5}=8\pi ^{2}/3$,...)
and a dot denotes differentiation with respect to the cosmic time $t.$ 
We may combine Eqs. (1), (4) and (5) to obtain an equation involving only the scale factor
$a(t)$. Furthermore, if we introduce a new dependent variable given by
$u_{D}\equiv\frac{a_{D}^{\prime }}{a_{D}}=\frac{dln a_{D}}{d\eta }$ 
and define conformal time as
$\eta=c\int_{}^{t}\frac{dt^{\prime }}{a(t^{\prime })}$ we get the Riccati equation
\begin{equation}
u_{D}^{\prime }+c_{D}u_{D}^{2}-c_{D}=0,  
\end{equation}
where
$c_{D}=\frac{D}{2}\gamma_{D}-1$
and a prime denotes differentiation with respect to $\eta$. In the following, we will consider
values of $\gamma_{D}$ such that $c_{D}\neq 0$. Eq. (6) can easily be solved by setting 
$u_{D}\equiv\frac{1}{c_{D}}\frac{w_{D}^{\prime }}{w_{D}}=\left[ \ln \left(
w_{D}^{1/c_{D}}\right) \right] ^{\prime }$
which gives the solution
\begin{equation}
a_{D}\left( \eta \right) = a_{0,D}\sinh \left( c_{D}\eta +\delta _{D}\right)
^{1/c_{D}},  
\end{equation}
where we have introduced the subscript D to identify the era which it 
applies and $a_{0,D}$ and $\delta_{D}$ are the integration constants to be 
determined from the initial conditions.

We also assume that at the end of the prematter era the fluid pervading the universe 
attained thermal equilibrium with a thermal spectrum. At this point we take the density 
of the universe equal to that of a massless scalar field 
with a thermal spectrum at $T_{pl,D}$, i.e.,
\begin{equation}
\rho \left( \eta _{D}\right) =\frac{A_{D}}{\left( 2\pi c\hbar
\right)^{D}}\Gamma \left( D+1\right) \xi\left( D+1\right)
\left( kT_{pl,D}\right) ^{D+1}.  
\end{equation}
Then, it is possible to find the duration  of the initial prematter era
by making use of Eqs. $(4)$ and $(8)$.

According to our consideration, dimensional reduction takes place via a
first order phase transition. But, its duration could be taken negligibly
small when compared with the durations of the constant $D$ eras in the
history of the universe. Thus, dimensional reduction could be considered to be
taking place at an instant of time. Denoting this time as $\eta _{D}$, we
may write Eq. (4) in terms of conformal time $\eta$ just before the dimensional 
reduction at $\eta=\eta _{D}-\mid \epsilon \mid $ as
\begin{equation}
\frac{D(D-1)}{2}\left[ \left( \frac{a_{D}^{\prime }}{a_{D}^{2}}\right) ^{2}-%
\frac{1}{a_{D}^{2}}\right] =\frac{2A_{D}G_{D}}{c^{4}}\rho _{D},
\end{equation}%
where $\epsilon $ is a small number. After the dimensional reduction at $%
\eta _{D},$ the number of space dimensions becomes $(D-1)$ and Eq. (9) takes 
the following form
\begin{equation}
\frac{(D-1)(D-2)}{2}\left[ \left( \frac{a_{D-1}^{\prime }}{a_{D-1}^{2}}%
\right) ^{2}-\frac{1}{a_{D-1}^{2}}\right] =\frac{2A_{D-1}G_{D-1}}{c^{4}}\rho
_{D-1},
\end{equation}%
at $\eta =\eta _{D}+\mid \epsilon \mid.$ In Eqs. (9) and (10), $G_{D}$ and $G_{D-1}$ denote the 
gravitational constant in $D$ and $D-1$ dimensions, respectively. At this point, as the most natural
and simple boundary conditions we take $a_{D}(\eta )$ and $a_{D}^{\prime
}(\eta )$ as continuous physical quantities at the transition points. Of
course, strictly speaking neither of them could be continuous. However,
since the phase transitions take place in a time interval whose duration is
expected to be significantly smaller than those of the constant $D$ eras and
the evolution of the universe during these eras is inflationary,
we could argue that during the transition period the change in $a_{D}(\eta)$ is
negligible. On the other hand, in the limit as $\epsilon \rightarrow 0$ a
discontinuity in $a_{D}^{\prime}(\eta)$ would imply infinite forces acting on the
universe, which in turn implies infinite values for the combinations $G_{D}\rho _{D}$ and $G_{D}P_{D}$
during  the dimensional reduction. Thus, for the time being we argue that taking 
$a_{D}^{\prime}(\eta )$ as continuous at the transition points is a good working
assumption. We now consider Eqs. (9) and (10) in the limit as $\epsilon \rightarrow 0$ 
and take their ratios to obtain
\begin{equation}
\frac{G_{D}}{G_{D-1}}=\frac{D}{D-2}\frac{A_{D-1}}{A_{D}}\left[ \frac{\rho
_{D-1}}{\rho _{D}}\right] _{\eta =\eta _{D}}.  \label{36}
\end{equation}%
With the help of Eq. (11) we relate the change in $%
G_{D} $ to that in $\rho _{D}.$ Thus, in order to find how\ $G_{D}$ changes
as $D$ changes, we have to know the discontinuity in density when the
universe experiences a dimensional reduction. We may express this discontinuity
by defining
\begin{equation}
l_{D,D-1}=\left[ \frac{\rho _{D-1}}{\rho _{D}}\right] _{\eta =\eta _{D}},
\label{39}
\end{equation}%
which gives us a "critical length parameter" with $[l_{D,D-1}]=cm$. Now, we 
could write the following expression for the change in $G_{D}$:
\begin{equation}
\frac{G_{D}}{G_{D-1}}=\frac{D}{D-2}\frac{A_{D-1}}{A_{D}}l_{D,D-1},
\end{equation}
where $l_{D,D-1}$ defines a characteristic length scale which marks the
point at which dimensional reduction occurs.

\section{Numerical Results and Conclusion}

In order to demonstrate the basic features of our model we now consider a numerical model 
which starts from a $D=4$ prematter era and evolves into a $D=3$ prematter. In this model,
to produce numerical results for the cosmological parameters of the universe one has to assign numerical 
values to the characteristic parameters of the model, i.e., $c_{3}$, $c_{4}$, $l_{4,3}$ 
and the initial value of the Hubble constant $H(0)$. It is apparent from Eqs. (1) and (7) that 
$c_{3}$ and $c_{4}$ represent the vacuum dominance of the universe in
$D=3$ and $D=4$ prematter eras, respectively. In determining the critical 
parameters of the model we first consider the present value of the density:
\begin{equation}
\rho (\eta _{now})=6.2269\cdot 10^{-35} T_{m}\left( \frac{4l_{4,3}}{\pi l_{pl}%
}\right) ^{\frac{2(2-c_{3})}{3(2c_{3}-1)}}gr/cm^{3},  \label{55}
\end{equation}
where $T_{m}$, which stands for the recombination temperature, is roughly at the order of $10^{3} K$ 
(see, e.g., Ref. 7, pp. 70-82). 
Furthermore, since the evolution of the universe is inflationary during the prematter eras, we have 
\begin{equation}
c_{4}\hspace{0.15cm}and\hspace{0.15cm}c_{3}\in \left( -1,0\right) .  
\end{equation}
It is evident from Eqs. (14) and (15) that as the numerical value of the term $(4l_{4,3}/\pi l_{pl})$
gets far from unity, the present value of the density predicted in Eq. (14) falls rapidly below 
$10^{-31}gr/cm^{3}$ which represents the order of the present value of the total density. Hence, we may 
conclude that a phase transition is signalled when the critical length parameter approaches the Planck 
length of the lower dimension. 

We also note that while our model is insensitive to the choice of $c_{3}$, its predictions strongly 
depend on the choice of $c_{4}$ and $H(0)$. We get numerical results that comply with recent 
observations\cite{Bennett}
only when $c_{4}$ gets numerical values close to -1 and $H(0)$ remains at the order of 
$10^{63}\hskip 0.1 cm km/sec.mpc$. As a specific example, we took $T_{m}=3000 \hskip 0.1 cm K$ and considered the following 
choices of parameters: $H(0)=1.6\cdot10^{63}\hskip 0.1 cm km/sec.mpc$ and
$l_{4,3}=1.4\cdot10^{-33}\hskip 0.1 cm cm$. For this particular case, it
is to be noted that only for a narrow range of initial conditions corresponding to the values of $c_{4}$ between 
$-0.99145$ and $-0.99141$, outcomes of this specific model are within the observed ranges for the cosmological
parameters, i.e., Hubble constant $H_{0}$ ($60\hskip 0.1 cm km/sec.mpc-80\hskip 0.1 cm km/sec.mpc$), 
age $t_{0}$ ($1.2\cdot10^{10}\hskip 0.1 cm yrs-1.6\cdot10^{10}\hskip 0.1 cm yrs$) and density 
$\rho_{0}$ ($\sim 10^{-31}gr/cm^{3}$).

\end{document}